\documentclass[aps,twocolumn,showpacs,preprintnumbers,prl,superscriptaddress]{revtex4}

\usepackage{graphicx}

\newcommand{\R}{\mathbf{r}}

\begin{document}

\title{Testing the broad applicability of the PBEint GGA functional and its one-parameter hybrid form}

\author{E. Fabiano}
\affiliation{National Nanotechnology Laboratory (NNL), Istituto Nanoscienze-CNR, via per Arnesano 16, I-73100 Lecce, Italy}

\author{Lucian A. Constantin}
\affiliation{Center for Biomolecular Nanotechnologies @UNILE, Istituto Italiano di Tecnologia, Via Barsanti, I-73010 Arnesano, Italy}

\author{F. Della Sala}
\affiliation{National Nanotechnology Laboratory (NNL), Istituto Nanoscienze-CNR, via per Arnesano 16, I-73100 Lecce, Italy}
\affiliation{Center for Biomolecular Nanotechnologies @UNILE, Istituto Italiano di Tecnologia, Via Barsanti, I-73010 Arnesano, Italy}

\date{\today}

\begin{abstract}
We review the performance of the PBEint GGA functional
(Phys. Rev. B 2010, 82, 113104) recently proposed to improve the
description of hybrid interfaces, and we introduce its 
one-parameter hybrid form (hPBEint).
We consider  different well established benchmarks for energetic 
and structural properties of molecular and solid-state
systems as well as model systems and newly developed 
benchmark sets for dipole moments and metal-molecule interactions.
We find that  PBEint and hPBEint (with 16.67\% Hartree-Fock exchange)
yield the overall best performance, working well for most 
of the considered properties and systems and showing a 
balanced behavior for different problems. In particular, due to
their well-balanced accuracy, they perform well for the description
of hybrid metal-molecule interfaces.
\end{abstract}

\pacs{71.10.Ca,71.15.Mb,71.45.Gm}

\maketitle

\renewcommand{\it}[1]{\underline{#1}}

\section{Introduction}
Density functional theory \cite{dftbook} (DFT) is nowadays one of the 
most popular computational methods in quantum chemistry and 
solid-state physics. Despite its success however, DFT still suffers
some limitations due the approximations used in the exchange-correlation
(XC) functional, which describes all the quantum 
electron-electron interactions. In fact, the development and 
optimization of advanced XC functionals is currently a very active 
field of research \cite{scuseria05}.

In the years numerous different XC functionals have been developed,
ranging from local density approximations \cite{ks} (LDA) to the 
most advanced orbital-dependent XC functionals 
\cite{kummel08,fdsexx,lhf,ls2,bartlett06} including exact exchange and 
selected exact-correlation contributions. The largest popularity
in practical applications is however own by two classes of functionals: 
generalized gradient  approximations (GGA) and hybrid functionals. 
The former are in fact, the method of choice for the investigation of 
large systems (e.g. in biochemistry or solid-state physics) 
due to their very favorable cost-to-accuracy ratio, while the latter
are the workhorse for computational chemistry. Moreover, the
GGA functionals attract theoretical interest because they are the building
blocks of the  more advanced meta-GGA, hybrid, and hyper-GGA functionals. 

Due to their simple form GGA functionals cannot fulfill all the 
exact constraints of the XC energy and cannot be accurate for
both atoms and solids \cite{perdew06,perdew08}. Therefore, some
selective criteria must be employed for their construction.
Popular approaches are to develop rather specialized functionals
by fitting to specific problems and training sets or by requiring
the satisfaction of selected exact constraints of the XC energy. 
As a result, GGA functionals for molecules (PBE \cite{pbe}, APBE \cite{apbe},
revPBE \cite{revpbe},RPBE \cite{rpbe}, BLYP \cite{b88,lyp}, 
OLYP \cite{lyp,optx}) or solids (PBEsol \cite{perdew08}, ARPA+ \cite{arpa},
AM05 \cite{am05}, Wu-Cohen \cite{wc}) are obtained. 
On the other hand, hybrid functionals rely strongly on the
underlying GGA approximations and share most of their fundamental
drawbacks. In addition, they might suffer from a certain
degree of empiricism used to construct the partial inclusion of the
Hartree-Fock exchange. Thus, hybrid functionals generally improve over 
the GGA description but cannot provide a really homogeneous level
of accuracy for a wide range of applications, showing occasionally
important failures for specific problems (e.g. transition-metal
chemistry \cite{furche06} or metal clusters \cite{pbeint_gold}).

Recently, a growing effort was devoted towards the development of
GGA functionals that could provide a well balanced description of 
a large number of properties. In fact, a wise selection
of important exact constrains of the XC energy (as e.g. in
PBE or APBE) can be used to optimize the performance of GGA
functionals in different contexts \cite{mukappa}. 
Two noteworthy examples are the PBEint \cite{pbeint} and the HTBS 
\cite{htbs} functionals.
These functionals are not only equally accurate for molecular and solid-state
energetic and structural properties, but potentially important
for interface physics, surface science and cluster chemistry.
Furthermore, they constitute an optimal starting point for the
construction of hybrid functionals of broad applicability.

In particular, the PBEint functional has been developed to 
respect different exact constraints of the XC energy 
and to connect the slowly-varying density regime, where the
second-order gradient expansion (GE2) of the exchange energy
is correct, and rapidly-varying density regime,
where the PBE behavior is accurate. Thus, it is 
reasonably accurate for an important set of molecular and solid-solid
state properties and, thanks to its well balanced behavior,
especially suited for interface \cite{pbeint}, metal-cluster
\cite{pbeint_gold} and surface \cite{js} problems.
We recall also that the PBEint XC hole density shows 
exact properties \cite{js}, beyond the PBE and PBEsol ones \cite{CPP},
related to the accurate analysis of metallic surfaces \cite{PCP}.

In this paper we propose a short review and a more detailed assessment
of the performance of PBEint for different problems in chemistry and 
physics. In addition, we use the PBEint functional to build
a global hybrid functional and investigate the results of this for different
amounts of the Hartree-Fock (HF) exchange mixing. We find that a rather 
small amount of HF mixing (only 16.67\%) gives the best results over a broad 
range of properties. We recall that global PBE and PBEsol hybrids are
instead commonly constructed using  25\% and 60\% of HF mixing 
\cite{csonka10}, respectively. Thus, the PBEint exchange 
construction is more compatible with the exact exchange, a result that 
was already proved in the case of jellium surfaces \cite{js}.

\section{Theory and computational details}
The PBEint functional is constructed considering an exchange functional
\begin{equation}\label{e1}
E_x = \int \rho(\R)\epsilon_x^\mathrm{unif}(\rho(\R))F_x(s(\R))d\R,
\end{equation}
with a PBE-like enhancement factor \cite{pbe}
\begin{equation}\label{e2}
F_x(s) = 1 + \kappa - \frac{\kappa}{1+\mu s^2/\kappa},
\end{equation}
where $\rho$ is the electron density, 
$\epsilon_x^\mathrm{unif} = -(3/4)(3/\pi)^{1/3}\rho^{1/3}$ is the exchange
energy per particle of the uniform electron gas, and
$s=|\nabla\rho|/(2(3\pi^2)^{1/3}\rho^{4/3})$ is the reduced gradient for
the exchange.
In Eq. (\ref{e2}) the value of the parameter $\kappa$ 
can be easily fixed to 0.804
by imposing that the Lieb-Oxford bound \cite{lieb81} for the exchange
energy holds locally \cite {pbe} (i.e. for the exchange energy 
density at each point in space). The value of the $\mu$ parameter
can be instead fixed by noting that Eq. (\ref{e1}) has, through second-order, 
the gradient expansion
\begin{equation}\label{e3}
E_x^{(GE2)} = \int \rho(\R)\epsilon_x^\mathrm{unif}(\rho(\R))[1+\mu s^2]d\R\ ,
\end{equation}
so that $\mu$ can be just identified with the second-order coefficient of 
the exchange gradient expansion. However, no unique value can be fixed
for $\mu$ with this condition. In fact, in the slowly-varying density 
limit ($s\rightarrow 0$) $\mu=\mu^{GE2}=10/81$ \cite{antoniewicz85}, 
while in the rapidly-varying density limit ($s\gg 1$)  
$\mu=\mu^\mathrm{PBE}=0.2195$ was found to be a good choice. Finally, from
the semiclassical-atom theory it was found $\mu=\mu^{MGE2}=0.26$  
\cite{apbe,elliott09}.
For the PBEint functional we therefore abandon the idea of
a constant $\mu$ and consider instead an $s$-dependent $\mu$ with
the following ansatz
\begin{equation}\label{e4}
\mu(s) =  \mu^{GE2} + (\mu^\mathrm{PBE} - \mu^{GE2})\frac{\alpha s^2}{1+\alpha s^2},
\end{equation}
with $\alpha=0.197$. Equation (\ref{e4}) assures that the following
minimal constraints are satisfied: i) in the slowly-varying
density limit $\mu\rightarrow\mu^{GE2}$; ii) in the
rapidly-varying density limit $\mu\rightarrow\mu^\mathrm{PBE}$;
(iii) the fourth-order term ($\propto s^4$) in the exchange gradient 
expansion vanishes.

Due to Eq. (\ref{e4}) the exchange enhancement factor of PBEint
varies smoothly between that of PBEsol (at small $s$ values) and
that of PBE (at larger $s$ values), granting a good description
of all possible density regimes and of many different systems from
molecules to solids. Note however that the PBEint functional 
(unlike HTBS) is not a simple interpolation between PBE and PBEsol,
but provides instead a physically meaningful (although interpolated)
value for the $\mu$ parameter at different density-regimes.
In fact, a similar construction proved to be useful also in the
case of noninteracting kinetic energy functionals \cite{kpbeint}.

For the correlation part, the PBEint functional utilizes again a PBE-like
expression \cite{pbe}
\begin{equation}\label{e5}
E_c = \int\rho(\R)\left[\epsilon_c^\mathrm{unif}(r_s(\R),\zeta(\R)) + H(r_s(\R),\zeta(\R),t(\R))\right]d\R,
\end{equation} 
with
\begin{equation}\label{e6}
H(r_s,\zeta,t) = \gamma\phi^3\ln\left(1 + \frac{\beta}{\gamma}\frac{t^2+At^4}{1+At^2+A^2t^4}\right),
\end{equation}
where $\rho=\rho_\uparrow+\rho_\downarrow$ is the total density,
$r_s=[(4\pi/3)\rho]^{1/3}$ is the local Seitz radius, 
$\zeta=(\rho_\uparrow-\rho_\downarrow)/\rho$ is the relative spin
polarization, $\phi=[(1+\zeta)^{2/3} + (1-\zeta)^{2/3}]/2$ is
a spin scaling factor, $\epsilon_c^\mathrm{unif}$ is the correlation
energy per particle of the uniform electron gas, $A$ is a function
of $ \epsilon_c^\mathrm{unif}$ and $\phi$ (see Ref. \onlinecite{pbe}),
$t=(3\pi)^{1/6}|\nabla\rho|/(4\phi\rho^{7/6})$ is the correlation
reduced gradient, and $\gamma=(1-\ln 2)/\pi^2$ is a parameter
fixed by uniform scaling to the high-density limit of the (spin-unpolarized)
correlation energy.
The parameter $\beta$, which is fixed to 0.066725 in the original PBE by the
second-order gradient expansion of the correlation energy, is obtained 
for PBEint by fitting to a jellium surface reference system: 
$\beta=\beta^\mathrm{PBEint}=0.052$. In fact, 
the PBEint correlation can reproduce with high accuracy the 
wave-vector analysis of the correlation surface energies of 
jellium slabs of different thickness and $r_s$ \cite{js}, outperforming 
other PBE-like GGA functionals. This result is important to 
prove the ability of PBEint to describe accurately systems where
different density regimes coexist.  The value of $\beta^\mathrm{PBEint}$
is also intermediate between those of PBE and PBEsol assuring
an accurate  compromise for a large palette of systems ranging from
atoms to solids \cite{js}.

\subsection{Hybrid functional}
To construct our hybrid functional we follow the adiabatic connection
scheme introduced in Ref. \onlinecite{perdew96} where the XC (hybrid)
functional is obtained as the result of the coupling-constant
integration
\begin{equation}\label{e7}
E_{xc}^\mathrm{hyb} = \int_0^1E_{xc}^\mathrm{hyb}(\lambda)d\lambda,
\end{equation}
and the following ansatz is used for the hybrid-functional
coupling-constant decomposition
\begin{equation}\label{e8}
E_{xc}^\mathrm{hyb}(\lambda) = E_{xc}^\mathrm{GGA}(\lambda) + (E_x^\mathrm{HF}-E_x^\mathrm{GGA})(1-\lambda)^{n-1},
\end{equation}
with $E_x^\mathrm{HF}$ being the Hartree-Fock exchange energy,
$E_x^\mathrm{GGA}$ being the exchange energy of a given GGA functional,
and $E_{xc}^\mathrm{GGA}(\lambda)$ its coupling-constant decomposition.
The parameter $n$ controls the balance between the nonlocal Hartree-Fock
and the local GGA exchange at different values of the coupling constant
and is related to perturbation theory considerations \cite{perdew96}.

If in Eq. (\ref{e8}) we use GGA=PBE and $n$=4, after performing the
integration (\ref{e7}), the PBE0 \cite{pbe0} functional (also
known as PBE1PBE) is obtained. In this work instead we consider
GGA=PBEint and we end up with the hPBEint functional
\begin{equation}\label{e9}
E_{xc}^\mathrm{hPBEint} = E_{xc}^\mathrm{PBEint} + \frac{1}{n}(E_x^\mathrm{HF}-E_x^\mathrm{PBEint})\ .
\end{equation}
In this case we do not fix the value of the parameter $n$, but we will 
consider $n$= 4 ,5, and 6 in order to assess the optimal value of this parameter on
a broad range of situations.
In fact, using a value of $n=4$ can be better to 
describe molecular systems (as in PBE0), while larger values of the 
parameter might be more appropriate for transition-metals, clusters
and interfaces.

\begin{table*}
\begin{center}
\caption{\label{tab1}Mean absolute errors for different tests as obtained 
from PBE, PBEint, PBEsol, PBE0, and hPBEint calculations with $n$=4, 5, 6. 
For each line the best result is indicated  by bold face, the worst is 
underlined . 
All energies are in kcal/mol (except for COH6). All bond lengths are in m\AA{}.
Vibrational frequencies are in cm$^{-1}$.}
\begin{ruledtabular}
\begin{tabular}{lrrrrrrr}
Test & PBE & PBEint & PBEsol & PBE0 & \multicolumn{3}{c}{hPBEint} \\
     &     &        &        &      & $n=4$ & $n=5$ & $n=6$ \\
\hline
\multicolumn{8}{c}{Organic molecules}\\
AE6  & 14.5 & 24.7 & \it{34.9} & \bf{5.4} & 12.4 & 14.3 & 15.7 \\
W4   & 10.7 & 15.5 & \it{21.5} & \bf{5.7} & 6.3 & 7.5 & 8.5 \\
OMRE   & \bf{6.8} & 8.2 & 12.0 & 9.1 & \it{12.8} & 11.7 & 11.1 \\
DC9    & 10.6 & 15.5 & \it{17.6} & \bf{10.2} & 13.6 & 13.9 & 14.2 \\
K9     & 7.51 & 9.09 & \it{10.59} &  \bf{3.96} &  4.90 & 5.62 & 6.17 \\
MGBL19 & 9 & \it{10} & \it{10} & \bf{6} & 9 & 8 & 7 \\
F38     & 56.8 & 65.4 & \it{65.9} & 51.7 & 53.2 & \bf{45.3} & 45.6 \\
          &          &     &     &     &     &     & \\
\hline
\multicolumn{8}{c}{Transition metals}\\
TM10 & 13.4 & 15.5 & 18.3 & \it{18.4} & 16.2 & 15.3 & \bf{12.0} \\
AUnAE$^a$ & \bf{0.3} & 2.2 & \it{4.4} & 3.4 & 2.5 & 1.8 & 1.3 \\
TMRE   & \bf{3.7} & 6.9 & 9.9 & \it{11.1} & \it{11.1} & 9.4 & 8.3 \\
AUnBL  & 78 & 31 & \bf{25} & \it{86} & 58 & 56 & 55 \\
          &          &     &     &     &     &     & \\
\hline
\multicolumn{8}{c}{Non-bonded interactions}\\
HB6  & \bf{0.4} & 0.5 & \it{1.7} & 0.5 & 0.7 & 0.7 & 0.6 \\
DI6  & \bf{0.4} & \bf{0.4} & \it{1.0} & \bf{0.4} & \bf{0.4} & \bf{0.4} & \bf{0.4}\\
          &          &     &     &     &     &     & \\
\hline
\multicolumn{8}{c}{Solid-state properties} \\
LC6    & \it{59}   & 32   & \bf{22} & - & - & - & - \\
COH6$^b$   & 0.15 & \bf{0.14} & \it{0.21} & - & - & - & -  \\
\end{tabular}
\end{ruledtabular}
\end{center}
\flushleft
a) atomization energy per atom. \\
b) eV/atom.
\end{table*}
\begin{table*}
\begin{center}
\caption{\label{tab_dip} Dipole moments (Debye) for different test molecules 
as computed with different DFT methods. The mean error (ME), mean absolute 
error (MAE) and mean absolute relative error (MARE) of each method with 
respect to reference values are reported in the last lines. For each row 
the best result is highlighted in bold style, the worst one is underlined.}
\begin{ruledtabular}
\begin{tabular}{lrrrrrrrrr}
System      & PBE & PBEint & PBEsol & PBE0 & \multicolumn{3}{c}{hPBEint} &  Ref.\\
        & & & & & $n=4$ & $n=5$ & $n=6$ & \\
\hline
CO       &  0.22 &  \it{0.24} &  0.23 &  0.10 &  \bf{0.11} &  0.14 &  0.15 &  0.11 \\
CFCl$_3$ &  0.35 &  \it{0.33} &  0.35 &  \bf{0.42} &  0.41 &  0.40 &  0.39 &  0.45 \\
furan    &  0.59 &  \it{0.58} &  0.59 &  \bf{0.66} &  0.65 &  0.63 &  0.62 &  0.66 \\
OCS      &  0.75 &  \bf{0.74} &  0.77 &  \it{0.80} &  0.79 &  0.78 &  0.77 &  0.71 \\
BF       &  1.02 &  \it{1.04} &  1.03 &  \bf{0.94} &  0.95 &  0.97 &  0.98 &  0.79 \\
CF$_2$O  &  \bf{1.12} & \it{1.13} & \bf{1.12} & \bf{1.12} & \it{1.13} & \it{1.13} & \it{1.13} & 0.95 \\
phenol &  \bf{1.25} &  1.26 &  1.26 &  1.28 &  \it{1.29} &  1.28 &  1.28 &  1.22 \\
HCOOH &  1.42 &  \bf{1.42} &  \bf{1.42} &  \it{1.47} &  \it{1.47} &  1.46 &  1.45 &  1.41 \\
CH$_3$SH &  1.58 &  \bf{1.60} &  1.61 &  1.62 &  \it{1.64} &  1.63 &  1.62 &  1.52 \\
HNCO &  \bf{1.99} &  \bf{1.99} &  2.01 &  \it{2.07} &  \it{2.07} &  2.06 &  2.05 &  1.61 \\
CH$_3$COOH &  \bf{1.72} &  \bf{1.72} &  1.73 &  \it{1.79} &  \it{1.79} &  1.77 &  1.77 &  1.70 \\
CH$_3$OCHO &  \bf{1.86} &  \bf{1.86} &  1.87 &  \it{1.89} &  \it{1.89} &  1.88 &  1.88 &  1.77 \\
HF &  \it{1.78} &  1.79 &  1.80 &  1.83 &  1.84 &  1.83 &  \bf{1.82} &  1.82 \\
H$_2$O &  \bf{1.87} &  1.88 &  1.89 &  1.92 &  \it{1.93} &  1.92 &  1.91 &  1.85 \\
CH$_3$Cl &  \it{1.82} &  1.83 &  1.84 &  1.89 &  1.90 &  1.89 &  \bf{1.88} &  1.87 \\
CH$_3$ONO &  2.26 &  2.27 &  \it{2.30} &  \bf{2.21} &  \bf{2.21} &  2.23 &  2.23 &  2.05 \\
pyridine &  \bf{2.22} &  2.23 &  2.24 &  \it{2.27} &  \it{2.27} &  2.26 &  2.26 &  2.19 \\
H$_2$CO &  \it{2.23} &  2.24 &  2.25 &  2.42 &  2.42 &  \bf{2.38} &  2.36 &  2.39 \\
ClCN &  \bf{2.96} &  \bf{2.96} &  2.98 &  \it{2.99} &  \it{2.99} &  \it{2.99} &  2.98 &  2.82 \\
CuH &  \it{2.31} &  2.33 &  2.28 &  2.85 &  \bf{2.87} &  2.76 &  2.69 &  2.97 \\
HCN &  2.93 &  2.94 &  \bf{2.95} &  3.03 &  \it{3.04} &  3.02 &  \bf{3.01} &  2.98 \\
CHOCH$_2$OH &  \it{2.40} &  \it{2.40} &  2.43 &  \bf{2.57} &  2.57 &  2.54 &  2.51 &  2.73 \\
CH$_3$NO$_2$ &  3.41 &  3.41 &  \bf{3.43} &  \it{3.61} &  \it{3.61} &  3.57 &  3.54 &  3.46 \\
LiCl &  6.90 &  \it{6.89} &  6.90 & \bf{7.06} &  \bf{7.06} &  7.03 &  7.01 &  7.23 \\
N6 & 16.94 & 16.95 & \it{17.03} & \bf{15.74} & 15.77 & 15.98 & 16.14 & 11.56 \\
 &  &  &  &  &  &  &  &  \\
ME &  \bf{0.20} &  0.21 &  0.22 &  \it{0.23} &  \it{0.23} &  \it{0.23} &  0.22 & \\
MAE &  0.35 &  0.35 &  \it{0.36} &  \bf{0.27} &  \bf{0.27} &  0.28 &  0.29 &   \\
MARE & 13.5\% & \it{14.4\%} & 14.4 & 8.0\% &  \bf{7.9\%} &  8.9\% &  9.8\% &  \\
Std. Dev. & 1.10 &  1.10 &  \it{1.11} &  \bf{0.83} &  0.84 &  0.88 &  0.92 & \\
\end{tabular}
\end{ruledtabular}
\end{center}
\end{table*}

\subsection{Computational details}
We tested the PBEint and hPBEint functionals for a series of molecular
and solid state properties including atomization energies, structural
properties, non-bonded interactions and reaction energies. 
The calculations were performed with the PBEint functional as well
as with the hPBEint functional with $n$=4, 5, and 6. Moreover,
calculations employing the PBE \cite{pbe}, PBEsol \cite{perdew08},
and PBE0 \cite{pbe0} functionals were also carried out for comparison.
In all calculations a def2-TZVPP basis set \cite{weigend03,weigend05}
was employed. For transition metals the core electrons were replaced with
effective core potentials (ECP) \cite{andrae90,dolg87,fuentalba83}.
All calculations were performed at the optimized ground-state geometry,
except for those on the DM25 and small interfaces test sets (see below).

All calculations on molecular species were performed with a development
version of the TURBOMOLE program package \cite{turbomole}.
Calculations on bulk solids were performed with the FHI-AIMS 
program \cite{aims1,aims2} using the light basis-set and a 
$18\times18\times18$ $k$-point grid. In this case, scalar
relativistic effects were accounted for by the zeroth-order
relativistic approximation (ZORA) \cite{zora}.

In more details, in the present work the following tests were considered:
\begin{itemize}
\item[] \textbf{AE6}: Atomization energies of SiH$_4$, SiO, S$_2$, C$_3$H$_4$, 
C$_2$H$_2$O$_2$, and C$_4$H$_8$; reference data were taken from Ref.
\onlinecite{lynch03}. Note that the AE6 test set was build to be 
representative for the results of the large Database/3 \cite{lynch03_2},
including 109 atomization energies.
\item[] \textbf{W4}: Atomization energies from the W4-08woMR test set of
Ref. \cite{goerig10}. It includes 83 atomization energies of organic molecules
selected from the original W4 test set \cite{karton08} excluding 
multi-reference cases.
\item[] \textbf{TM10}: Atomization energies of CrH, MnH, CoH, V$_2$, 
Sc$_2$, TiO, MnO, CuO, CrF, and CuF. Reference data were taken from Ref.
\onlinecite{furche06}.
\item[] \textbf{AUnAE}: Atomization energies of the Au$_2^-$, Au$_2$, 
Au$_3$, and Au$_5$ clusters. Reference data, including relativistic and 
thermal corrections, were taken from Ref. \onlinecite{pbeint_gold}.
\item[] \textbf{K9}: Barrier heights and reaction energies of 
three organic reactions. Namely, OH+CH$_4\rightarrow$CH$_3$+H$_2$O,
H+OH$\rightarrow$O+H$_2$, and H+H$_2$S$\rightarrow$H$_2$+HS.
Reference data were obtained from Refs. \onlinecite{lynch03,lynch03_e}.
\item[] \textbf{HB6}: Binding energies of the hydrogen-bond interacting
systems (H$_2$O)$_2$, (HF)$_2$, (NH$_3$)$_2$, NH$_3$--H$_2$O,
(HCONH$_2$)$_2$, and (HCOOH)$_2$. Reference data were taken from
Ref. \onlinecite{zhao05}.  
\item[] \textbf{DI6}: Binding energies of the dipole-dipole interacting
systems CH$_3$Cl--HCl, CH$_3$SH--HCl, CH$_3$SH--NCH, (H$_2$S)$_2$, 
(HCl)$_2$, and H$_2$S-HCl. Reference data were taken from
Ref. \onlinecite{zhao05}.  
\item[] \textbf{MGBL19}: Bond lengths of H$_2$, CH$_4$, NH$_3$, H$_2$O, HF,
C$_2$H$_2$, HCN, H$_2$CO, OH, CO, N$_2$, F$_2$, CO$_2$, N$_2$O, and Cl$_2$.
Reference values were taken from Ref. \onlinecite{zhao06}. Note that this
set provides a global assessment over bond lengths involving one hydrogen
atom and bond lengths not involving hydrogen atoms, which usually behave
differently for different functionals \cite{mukappa}.
\item[] \textbf{AUnBL}: Bond lengths of Au$_2^-$, Au$_2$, Au$_3^+$,
Au$_3^-$, Au$_4$, Au$_6$ (capped pentagon, C$_{5v}$ symmetry),
SeAu$_2$, and (ClAuPH$_3$)$_2$. Reference data were taken from
Ref. \onlinecite{pbeint_gold}.   
\item[] \textbf{F38}: Harmonic vibrational frequencies of H$_2$, CH$_4$,
NH$_3$, H$_2$O, HF, CO, N$_2$, F$_2$, C$_2$H$_2$, HCN, H$_2$CO, CO$_2$,
N$_2$O, Cl$_2$, and OH. Reference data were taken from Ref.
\onlinecite{biczysko10}   
\item[] \textbf{OMRE}: Reaction energies of organic molecules. It includes
$\mathrm{CH}_2+\mathrm{H}_2\rightarrow\mathrm{CH}_4$,
$\mathrm{F}_2+\mathrm{H}_2\rightarrow 2\mathrm{HF}$,
$\mathrm{C}_6\mathrm{H}_6+3\mathrm{H}_2\rightarrow\mathrm{C}_6\mathrm{H}_{12}$,
$\mathrm{CO}+3\mathrm{H}_2\rightarrow\mathrm{CH}_4+\mathrm{H}_2\mathrm{O}$,
$\mathrm{SO}_2+3\mathrm{F}_2\rightarrow\mathrm{SF}_6+\mathrm{O}_2$, and
$\mathrm{C}_4\mathrm{H}_6+\mathrm{C}_2\mathrm{H}_4\rightarrow\mathrm{C}_6\mathrm{H}_{10}$.
Reference data were taken from Ref. \onlinecite{grimme06}
\item[] \textbf{TMRE}: Reaction energies of transition-metal complexes.
It includes $\mathrm{Ni(CO)}_3+\mathrm{CO}\rightarrow\mathrm{Ni(CO)}_4$,
$\mathrm{Fe(CO)}_4+\mathrm{CO}\rightarrow\mathrm{Fe(CO)}_5$,
$\frac{1}{2}\mathrm{Cl}_2+\mathrm{CoCl}_2\rightarrow\mathrm{CoCl}_3$,
and $2\mathrm{FeCl}_2\rightarrow\mathrm{Fe}_2\mathrm{Cl}_4$.
Reference data were taken from Ref. \onlinecite{furche06}.
\item[] \textbf{DC9}: Nine reaction energies of medium-size molecules,
which are usually treated poorly by DFT methods. Reference values
were taken from Ref. \onlinecite{goerig10}. 
\item[] \textbf{LC6} and \textbf{COH6}: Lattice constants and
cohesive energies of bulk Na (simple metal), Ag, Cu (transition metals), 
Si, GaAs (semiconductors), and NaCl (ionic solid). Reference data were
taken from Refs. \onlinecite{csonka09,haas09}. The LC6 and COH6 tests
were not performed for hybrid functionals.  
\item[] \textbf{DM25}: Dipole moments of molecules. The test set
considers a broad range of systems and of dipole moments, ranging from 
0.11 to 11.56 Debye. In detail, it includes 
CO, CFCl$_3$, furan, OCS, BF, CF$_2$O, phenol, HCOOH,
CH$_3$SH, HNCO, CH$_3$COOH, CH$_3$OCHO, HF, H$_2$O, CH$_3$Cl,
CH$_3$ONO, pyridine, H2CO, ClCN, CuH, HCN, CHOCH$_2$OH,
CH$_3$NO$_2$, LiCl, and NH$_2$(CH=CH)$_6$NO$_2$ (denoted as
N6). Reference data and geometries of
BF, H$_2$O, CuH, H$_2$CO, LiCl, and N6 are taken from Ref.
\onlinecite{zhao06_2}. The remaining reference data and 
(experimental) geometries are taken from Ref. \onlinecite{nist}.
All structures are available in supporting information \cite{suppinfo}.
\item[] \textbf{Small interfaces}: Interaction energies of twelve small
metal-molecule systems, representative of metal-molecule interfaces.
The set includes Au$_2$--SH, Au$_2^+$--N$_2$, 
Au$_2$--SHCH$_3$, Au$_3$--SH, Au$_3^+$--N$_2$,
Au$_3$--SCH$_3$, Au$_4$--SH, Au$_4^+$--N$_2$, and
Au$_4$--SHCH$_3$. The general structure of the molecule-cluster
systems was obtained from Refs. 
\onlinecite{kruger01,ding05,kuang11}. Each structure
was then reoptimized at the TPSS/def2-TZVPP
\cite{weigend03,weigend05,tpss} level of theory. 
This geometry was successively employed in all the calculations. 
Reference values were obtained from CCSD(T) calculations 
\cite{purvis82,scuseria88,pople87} extrapolated to the complete basis set
limit \cite{truhlar98}.
All structures are available in supporting information \cite{suppinfo}.
\item[] \textbf{Model systems}: We consider four sets of model systems
for which exact or extremely accurate solutions are available. Namely, we
consider: jellium surfaces, jellium clusters, 
Hooke's atoms, and the hydrogen atom with
fractional spin. Additional details are provided in the corresponding 
subsection later on.
\end{itemize}

All tests are necessarily not exhaustive, because for computational
reasons they only include a relatively small number of systems.
Therefore, for specific cases they might be not fully representative
of the true performance of the functionals. Nevertheless, since
each test set was constructed to have good representativity and
because the selected tests cover a broad range of properties and systems,
we believe that the present assessment offers a fair overview of 
the performance of the considered functionals.

\begin{table*}
\begin{center}
\caption{\label{tab2} Interaction energies (eV) of small organic molecules 
and gold clusters, computed with different DFT methods. The mean error (ME), mean 
absolute error (MAE) and mean absolute relative error (MARE) of each method with 
respect to CCSD(T) are reported in the last lines. For each row the result in best 
agreement with CCSD(T) calculations is highlighted in bold style, the one in worst 
agreement is underlined.}
\begin{ruledtabular}
\begin{tabular}{lrrrrrrrrr}
System      & PBE & PBEint & PBEsol & PBE0 & \multicolumn{3}{c}{hPBEint} & CCSD(T)\\
\cline{6-8}
        & & & & & $n=4$ & $n=5$ & $n=6$ & \\
\hline
Au$_2$--SH    & \bf{1.99} & 2.20 & 2.35 & \it{1.57} & 1.72 & 1.81 & 1.87 & 1.98 \\
Au$_2^+$--N$_2$&0.99 & 1.13 & \bf{1.27} & \it{0.84} & 0.95 & 0.99 & 1.01 & 1.30 \\
Au$_2$--SHCH$_3$&0.88& 1.04 & \bf{1.17} & \it{0.78} & 0.90 & 0.92 & 0.94 & 1.24 \\
Au$_3$--SH    & 3.02 & 3.20 & 3.33 & \it{2.96} & 3.09 & 3.11 & \bf{3.12} & 3.15 \\
Au$_3$--N$_2$ &0.77 & 0.93 & \it{1.06} & 0.67 & \bf{0.78} & 0.74 & 0.77 & 0.78 \\
Au$_3$--SCH$_3$&2.73 & \bf{2.90} & \it{3.03} & 2.66 & 2.79 & 2.81 & 2.82 & 2.88 \\
Au$_4$--SH    & \it{2.18} & 2.36 & \bf{2.49} & 2.09 & 2.23 & 2.25 & 2.27 & 2.48 \\
Au$_4^+$--N$_2$&0.67 & 0.80 & \it{0.92} & 0.58 & 0.68 & 0.70 & \bf{0.72} & 0.74 \\
Au$_4$--SHCH$_3$&0.95& 1.11 & \it{1.24} & 0.93 & \bf{1.04} & \bf{1.06} & 1.07 & 1.05 \\
              &      &      &  &      &      &      &      &      & \\
ME      & -0.16 & \bf{0.01} & 0.14 & \it{-0.28} & -0.16 & -0.13 & -0.11 & \\
MAE     & 0.16 &  \bf{0.12} & 0.16 & \it{0.28} & 0.16 & 0.14 & \bf{0.12} & \\
MARE    & 10.61\% & 8.84\% & 13.08\% & \it{18.89\%} & 10.30\% & 9.16\% & \bf{7.72\%} & \\
Std Dev & 0.14 & 0.14 & \it{0.15} & \it{0.15} & 0.14 & 0.13 & \bf{0.12} & \\
\end{tabular}
\end{ruledtabular}
\end{center}
\end{table*}

\section{Results}

\subsection{Quantum chemistry and solid-state benchmarks}

In Table \ref{tab1} we report the mean absolute errors (MAE) 
for different tests,
as obtained from PBEint and hPBEint calculations with $n=3,4,5$, with the 
aim of assessing the performance of the functionals for a broad set of 
problems. To this end, also PBE, PBEsol, and PBE0 results are reported,
since these are natural references for the functionals considered here.
Full results of all tests are available in supporting information
\cite{suppinfo}.

For atomization energies of organic molecules, it is well known that 
GGAs are not very accurate (PBEsol largely fails), while  the hybrid 
functionals perform better than the corresponding GGAs and this is more so
for those functionals including a larger fraction of Hartree-Fock exchange 
\cite{csonka10}. 
Consequently, PBE0, with a MAE of 5 kcal/mol, is the best functional overall
and hPBEint with $n=4$ yields the best performance among the different 
variants of hPBEint, reaching almost the PBE0 accuracy for the W4 test.
Note that even better results can be obtained by considering
higher fractions of exact exchange, as shown for example in
Ref. \onlinecite{csonka10} where the admixture of PBE GGA and 32\% of
exact exchange was found to yield the best results for thermochemistry.

Whereas hybrid functionals improve significantly
atomization energies, for reaction energies involving organic 
molecules the use of hybrid functionals is not so systematically
beneficial \cite{goerig10,goerig11,goerig11_2}.
In fact, inclusion of Hartree-Fock exchange brings some improvements 
in some cases, e.g. for the K9 test where PBE0 and hPBEint
with $n=4$ perform best (see also the BH76RC test \cite{csonka10,goerig10}),
but has small effects for other cases, e.g. G2RC, NBPRC and DC9 tests
\cite{goerig10,goerig11}.  
Furthermore, in most situations improvements are smaller or 
even absent when large fractions of Hartree-Fock exchange are used; 
see for example BHLYP vs BLYP and B3LYP for the BH76RC test 
\cite{csonka10,goerig10}.
In particular, for the DC9 and OMRE benchmarks considered here,
we find that PBE0 brings almost no improvement or even 
a worsening of the results with respect to PBE. Analogously,
the best results at the hPBEint level are obtained when $n=6$ and a
relatively small fraction of nonlocal exchange is used.

Different conclusions are found when one considers the energetic properties 
of transition-metal complexes or clusters (TM10, AUnAE, TMRE). In this case
in fact the use of PBE0 corresponds always to a considerable worsening 
of the results with respect to PBE. A worsening of the performance
is also observed in all cases for hPBEint with respect to PBEint when
$n=4$, although not so marked as for PBE0 vs PBE. Improvements with
respect to PBEint are instead found for hPBEint with $n=6$,
which becomes the best functional for TM10, but not for the TMRE test.
In this last case hPBEint with $n=6$ is definitely the best hybrid
functional in the present investigation, but slightly worst than PBEint and
more than twice worst than PBE. Note however that the TMRE test is a
particular difficult test in the context of the present work, because
it requires a delicate balancing between the description of
metal complexes (best described by PBEint and PBE) and
small organic molecules (best described by hybrids with a large fraction
of Hartree-Fock exchange).

Concerning structural properties, hybrid functionals provide a modest 
improvement for organic molecules (MGBL19 and F38), while they
worsen the results for gold clusters (AUnBL). Interestingly, the hPBEint
functional with $n=6$ yields the best results among the different
variants of hPBEint, outperforms PBE0 for AUnBL and F38, and
gives almost the same result as PBE0 for MGBL19. Thus,
the combination of PBEint with 1/6 of Hartree-Fock exchange proves
to be a very good choice for structural properties of finite systems.

\begin{table*}
\begin{center}
\caption{\label{tab_model} Mean absolute errors (MAE) in Hartree or 
mean absolute relative errors (MARE) for different functionals for 
the proposed model systems. Best results are in bold style; worst results are underlined.}
\begin{ruledtabular}
\begin{tabular}{lrrrrrrrr}
System      & PBE & PBEint & PBEsol & PBE0 & \multicolumn{3}{c}{hPBEint} \\
        & & & & & $n=4$ & $n=5$ & $n=6$ \\
\hline
Jellium surf. energy (MARE)  & \it{3.15} & 2.29 & 2.32 & \bf{1.88} & 2.29 & 2.29 & 2.29 \\
Jellium spheres (MAE)        & 0.46 & 0.46 & 0.56 & \it{0.57} & 0.23 & 0.18 & \bf{0.17} \\
Hooke's atom (MARE)          & 3.63 & 3.45 & \bf{3.41} & \it{3.87} & 3.78 & 3.71 & 3.66 \\
Fractional spin H (MAE)      & 29.59 & 27.78 & \bf{27.63} & \it{42.31} & 40.80 & 38.19 & 36.46\\
\end{tabular}
\end{ruledtabular}
\end{center}
\end{table*}

For non-bonded interactions very good results are obtained
at the GGA level using the PBE and PBEint functionals, while
poorer results (especially for hydrogen bonds) are given by the
PBEsol functional. Hybrid functionals cannot improve the 
performance of PBE and PBEint, yielding the same MAE as the GGAs
for the DI6 test and slightly larger errors for the HB6 test.
In this latter case, the best results at the hybrid level
are obtained by PBE0, taking advantage of the high accuracy of
PBE for the hydrogen bonds, and by hPBEint with $n$=6,
taking advantage of the small fraction of Hartree-Fock exchange
included.

Concerning solid-state properties, PBEsol is the most accurate for
lattice constants, while PBEint is intermediated between it and PBE.
However, the limited accuracy of PBEsol for atomic energies
causes it to perform poorly for cohesive energies (COH6 test).
In this case, where a good description of both bulk solids 
and atoms is required, PBEint shows a remarkably good
performance, due to its well balanced description of different
density regimes.

\subsection{Dipole moments}
In Tab. \ref{tab_dip} we report the performance of the different functionals
for the description of dipole moments of several test molecules.
Benchmarking the dipole moments is not as common as energy and 
structural tests, nevertheless it is very
important to verify the accuracy in the description of density 
\cite{laricchia10,ijqc}.
This test allows in fact to probe indirectly the quality of the description,
provided by different functionals, of the electron density distribution in 
molecules, especially in valence and asymptotic regions.
This issue is particularly relevant for GGA functionals, because due to 
the Coulomb self-interaction error, often they yield a poor
description of the electron density far from the atomic core.
For the same reasons, an improved description is usually obtained by 
hybrid approaches.

The general trends outlined above are confirmed by inspection of the
results of Tab. \ref{tab_dip}. The smallest errors are obtained
by the hybrid functionals, which yield a global MAE of about
0.28 Debye. A MAE larger by 30\% is found instead for GGA methods 
which perform all very similarly and
display a global mean absolute error of about 0.35 Debye.
In particular, the best results are obtained when the larger
fraction of Hartree-Fock exchange (0.25 for PBE0 and hPBEint
with $n$=4) is considered, having a MAE of 0.27 Debye and
a MARE below 8\%. Nevertheless, good results are achieved
also by hPBEint with $n$=6, including only a relatively
small fraction of Hartree-Fock exchange.

We note however that the results of Tab. \ref{tab_dip} show
for all the functionals several important deviations from the
trends depicted above. 
In fact, for some systems (e.g. HNCO, pyridine, ClCN)
the best results are achieved at the GGA level, while
PBE0 yields the worst agreement with the reference.
Thus, none of the functionals considered here can be 
labeled as fully reliable for the calculation
of the dipole moment of an arbitrary molecular system, but
the best compromises are provided by the hPBEint
functional with $n$=6, which yields the largest
number of results with errors $\leq$0.1 Debye,
and PBE0 that yields the best overall MAE and MARE and
the largest number of results with errors
$\leq$0.2 Debye.

\begin{table*}
\begin{center}
\caption{XC jellium surface energies ($\mathrm{erg}/\mathrm{cm}^{2}$) of 
the global hybrids PBE0 and hPBEint. The fixed-node diffusion Monte Carlo 
(DMC) calculations \cite{WHFGG,WHFGG2} are reported for reference.
($1\mathrm{hartree}/\mathrm{bohr}^2=1.557\times 10^6
\mathrm{erg}/\mathrm{cm}^{2}$.). PBE, PBEint, and PBEsol results are 
taken form Table I of Ref. \onlinecite{pbeint}.}
\begin{ruledtabular}
\begin{tabular}{lrrrrrrrrr}
$r_s$   & PBE & PBEint & PBEsol  & PBE0 & \multicolumn{3}{c}{hPBEint} &  DMC\\
        &     &        &         & & $n=4$ & $n=5$ & $n=6$ & \\
\hline
2 & 3265 & 3378 & 3374 & 3312 & 3378 & 3378 & 3378 & 3392 $\pm$ 50 \\
3 & 741 & 774 & 774 & 756 & 774 & 774 & 774 & 768 $\pm$ 10 \\
4 & 252 & 267 & 267 & 259 & 267 & 267 & 267 & 261 $\pm$ 8 \\
6 & 52 & 56 & 56 & 55 & 56 & 56 & 56 & 53 $\pm ...$ \\
\end{tabular}
\end{ruledtabular}
\label{tab3}
\end{center}
\end{table*}

\subsection{Metal-molecule interfaces}

In this subsection we investigate the ability of different functionals
to describe hybrid metal-molecule interfaces. These systems pose 
a difficult challenge to any computational method because of the very 
different theoretical and numerical issues raised by metallic (extended)
systems and molecules. The former are in fact mainly characterized by 
a slowly-varying density regime and can be properly described with local 
or semilocal approaches. The latter instead display a significant
contribution from rapidly-varying density regions and require
a proper treatment of nonlocal interactions.

In view of practical applications it would be nice, of course, to perform 
tests for the interaction of (functionalized) organic molecules with 
relatively large metallic clusters or extended (eventually semi-infinite) 
surfaces.
However, for these systems no benchmark value to assess the performance
of the methods can be obtained. Accurate experimental values are in 
fact extremely rare, while high level approaches cannot be applied due
to their exceedingly high computational cost. Note also that for large
metallic systems static and high-order correlation effects start to
play a fundamental role, thus accurate results can only be obtained
through very sophisticated methods.

For the reasons mentioned above, in this paper we restrict our attention
to a set of relatively small metal-molecule systems composed of 
small organic molecules interacting with gold clusters of 2, 3 and 4
atoms. For such systems quite accurate benchmark results have been
obtained with coupled cluster single, doubles with perturbative triple
correction (CCSD(T)) calculations and an assessment of DFT methods
could be carried out.

Results are reported in Tab. \ref{tab2}.
At the GGA level, PBE presents a general tendency to underestimate the
interaction energies, while a slight overestimation is observed by PBEint.
Finally, PBEsol shows a rather strong 
overestimation of the interaction energies.
Overall, PBEint (with a MAE of 0.16 kcal/mol) is slightly better than 
PBE and PBEsol.
In addition, we note that most of the accuracy for
PBE comes from the smallest systems (e.g., Au$_2$--SH)
and larger errors are found when the system size increases. For PBEint an 
opposite behavior is observed. This trend was already observed for gold
clusters \cite{pbeint_gold} and led to the conclusion that PBEint 
provides the most accurate description of medium and large metal clusters. 
A similar extrapolation can be supposed to apply also in the present
case of metal-molecule interfaces, suggesting that superior results can
probably be achieved by PBEint for realistic interfaces. This conclusion is
also supported by the observation that for interfaces involving large 
systems slowly-varying density regions will gain a more important role.
Thus PBEint (and partially PBEsol) will be favored over PBE.

The addition of a fraction of Hartree-Fock exchange in the functionals
has the effect to reduce the computed interaction energies. This results
in a worsening of PBE0 with respect to PBE, but an improvement of
hPBEint with respect to PBEint. In fact, all the variants of
hPBEint show a MARE lower than PBE and PBEint. In particular, the
option with $n=6$ displays the smallest MAE and MARE of all the 
functionals considered in this work. We note also that,
because of the small fraction of Hartree-Fock exchange that it includes,
hPBEint with $n=6$ may be supposed to be the best compromise
for interfaces of interest in practical application, where large metallic 
systems are involved \cite{jcp_interface,vonc}.

\subsection{Model systems}
In this subsection, we show the performance of PBE0 and hPBEint for 
several important model-systems: 

(i) jellium surfaces, that contain
the main physics of simple metal real surfaces; 

(ii) jellium spheres, that are simple models for 
simple-metal-particles;  

(iii) Hooke's atom,
that represents two interacting electrons
in an isotropic harmonic potential of frequency $\omega$.
At small values of $\omega$, the electrons are strongly correlated, and
at large values of $\omega$, they are tightly bounded, two important cases
in many condensed matter applications.  

(iv) The hydrogen atom with fractional spin, a model 
for the static correlation given by degenerate states.
 
For all these cases, we use non-self-consistent accurate calculations
(for jellium surfaces we use numerical LDA orbitals and densities,
for jellium spheres we use accurate exact exchange orbitals and densities, 
for the Hooke's atom we use exact orbitals and densities, and for 
the spin-dependent hydrogen atom we use
exact-exchange orbitals and densities.) For the hybrids, 
we use exact-exchange instead of the Hartree-Fock one.

The mean absolute (relative) errors of different functionals for the
selected model problems are summarized in Tab. \ref{tab_model}.
A detailed analysis of the results is given in the following.

\subsubsection{Jellium surfaces}
The PBEint GGA performs remarkably well for the jellium surfaces,
giving very accurate exchange and XC surface energies \cite{pbeint,js}. 
Because there are 
no error compensations between the exchange and correlation parts
and the  PBEint exchange surface energies are extremely close
to the exact-exchange ones \cite{js},
any PBEint hybrid will yield the same results as the PBEint GGA and 
thus be very accurate for the jellium surfaces, as 
we explicitly show in the Table \ref{tab3}. This is a very important 
result for a functional designed for hybrid interfaces, where the 
surface physics plays a dominant role. 
We also mention that PBE0 
improves considerably over PBE, but still it is not so accurate 
in the range $2\leq r_s \leq 3$, where important metals lie
(e.g. Al has $r_s=2.07$, and Cu has $r_s=2.67$).  

We mention that the RGE2 GGA of Ref. \cite{RGE2} gives similar jellium 
xc and x-only surface energies as PBEint (see Table 3 and Fig. 3 of 
Ref. \cite{RGE2} for a detailed analysis of jellium surfaces). Both RGE2 and 
PBEint have the exchange enhancement factor $F_x\rightarrow 1+\mu s^2+O(s^6)$
at small $s$, and thus the requirement of vanishing $O(s^4)$ terms in this 
expansion seems very important for accurate jellium x-only surface energies
at the GGA level. Note that the exchange part of revTPSS meta-GGA 
\cite{revTPSS}, 
which recovers the 4-th order gradient expansion of the exchange energy, 
is also accurate for jellium surfaces.

\subsubsection{Jellium spheres}
\begin{figure}
\includegraphics[width=\columnwidth]{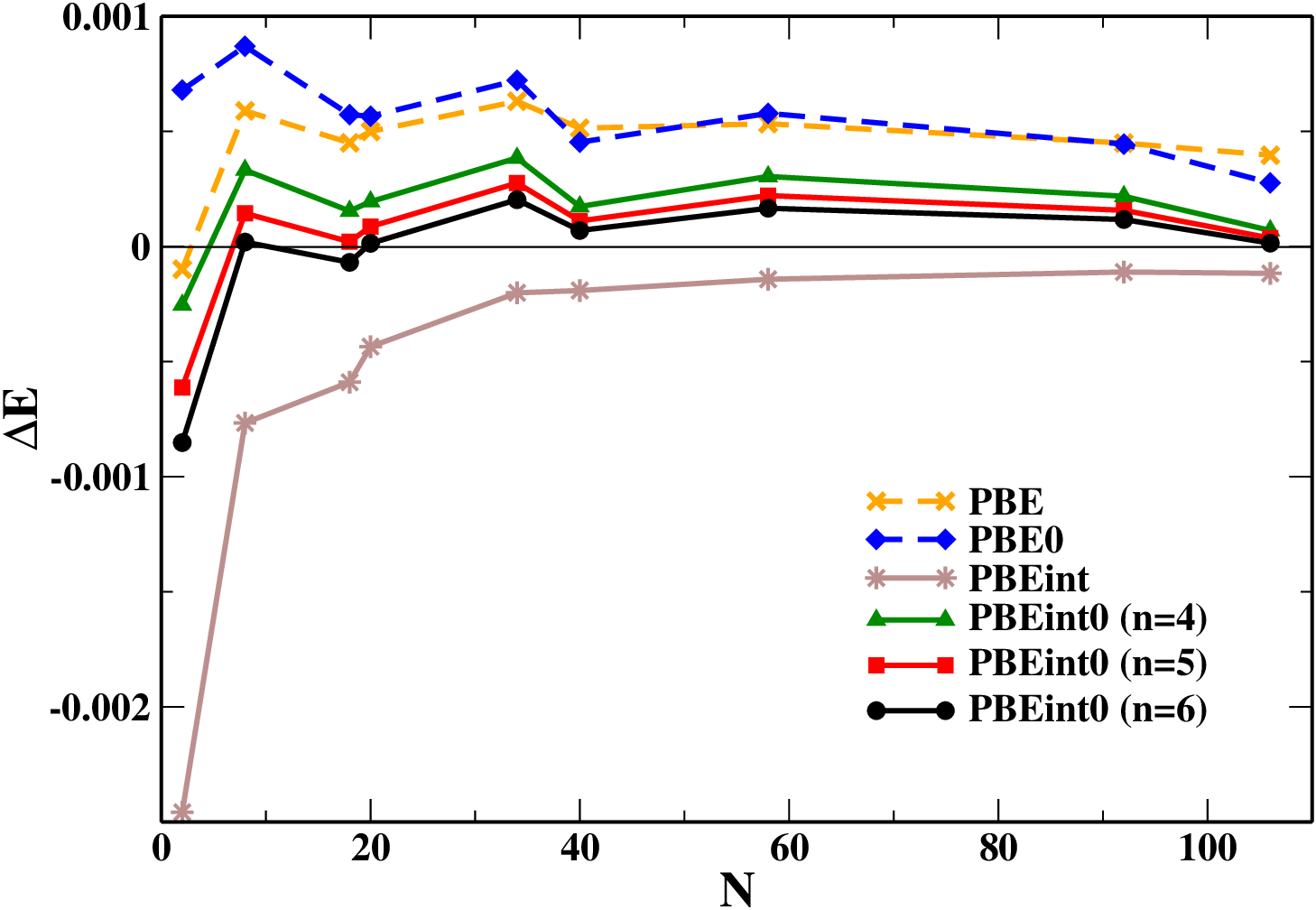}
\caption{Errors in total energies per electron ($E^{exact}/N-E^{approx}/N$; 
Hartree) of jellium spheres for $r_s=4$. $E^{exact}/N$ represents the corrected 
DMC data \cite{TPAFK}. We use magic neutral clusters with 
$N$=2$e^-$, 8$e^-$, 18$e^-$, 20$e^-$, 34$e^-$, 40$e^-$, 58$e^-$, 92$e^-$, and 106$e^-$.}
\label{fjs}
\end{figure}

In Fig. \ref{fjs}, we show the errors in total energies per 
electron ($E^{exact}/N-E^{approx}/N$) versus the number 
of electrons $N$, for jellium spheres of magic numbers
($N$=2, 8, 18, 20, 34, 40, 58, 92, and 106) with bulk parameter $r_s=4$. 
(Note that sodium has $r_s=3.93$.) The reference values $E^{exact}/N$ 
represent the corrected DMC data (See Eq. (40) and Fig. 5 of Ref. \onlinecite{TPAFK}).
We note that HF and EXX perform similarly and badly for this problem \cite{Sottile},
whereas LDA is remarkably accurate for large clusters \cite{CR1,LCPB}. 
We also recall that self-consistent effects are very small (see Table I of \cite{APF}), 
such that our non-self-consistent results are expected to give an honest and 
accurate picture of the performance of various functionals for 
jellium (metal) particles.
%
%

We observe that PBE0 does not improve upon PBE, being even worse for 
$N\leq 34$. On the other hand, hPBEint (with n=4, 5 and 6) significantly 
improve over the original PBEint for $N\leq 20$. However, on average the best 
performance is given by hPBEint (n=6). This is also a very important
result for a functional designed for hybrid interfaces, where metal particles 
play a crucial role.

\subsubsection{Hooke's atom}
\begin{figure}
\includegraphics[width=\columnwidth]{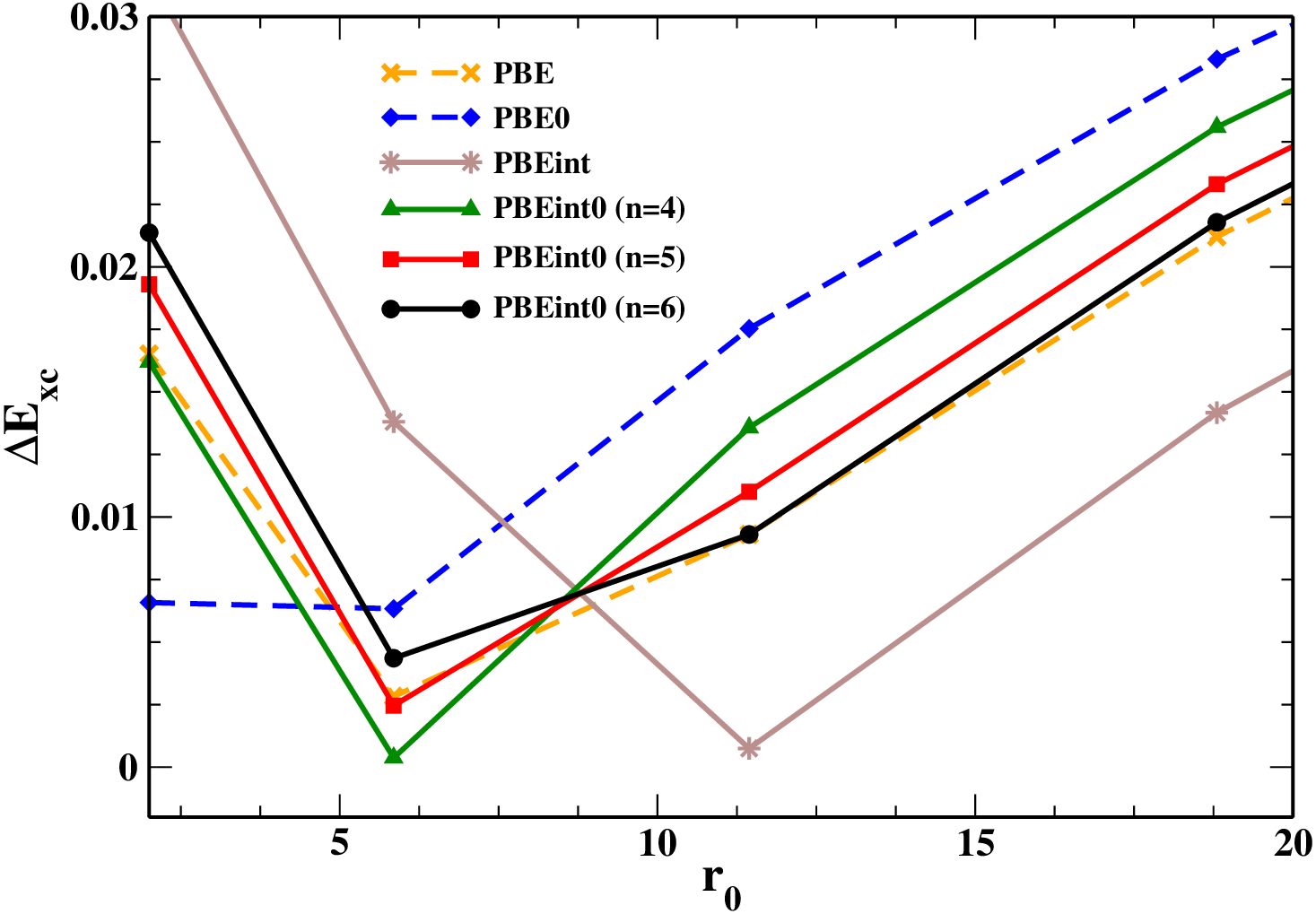}
\caption{ $\Delta E_{xc}=|(E_{xc}^{app}-E_{xc}^{exact})/E_{xc}^{exact}|$ 
(Hartree) for 
the Hooke's atom with different frequencies. $r_0=(\omega^2/2)^{-1/3}$
is the classical electron distance \cite{js}.}
\label{f1}
\end{figure}

In Fig. \ref{f1}, we report the relative absolute errors of the 
hybrids, for the Hooke's atom with several frequencies $\omega$.
In the tightly-bounded regime ($r_0=(\omega^2/2)^{-1/3}$ is small)
all the functionals performs similarly, though PBE0 
is the best in this extreme case. On the other hand, for $r_0 >5$, that is the 
physical case for most of the real applications, the hPBEint hybrids 
are always better than PBE0, with a superior performance for the
$n=6$ case.
However, the errors increase when the strong correlation increases,
showing that these global hybrids can not describe satisfactorily 
the strongly correlated regime.  
%
%

\subsubsection{Hydrogen atom with fractional spin}

\begin{figure}
\includegraphics[width=\columnwidth]{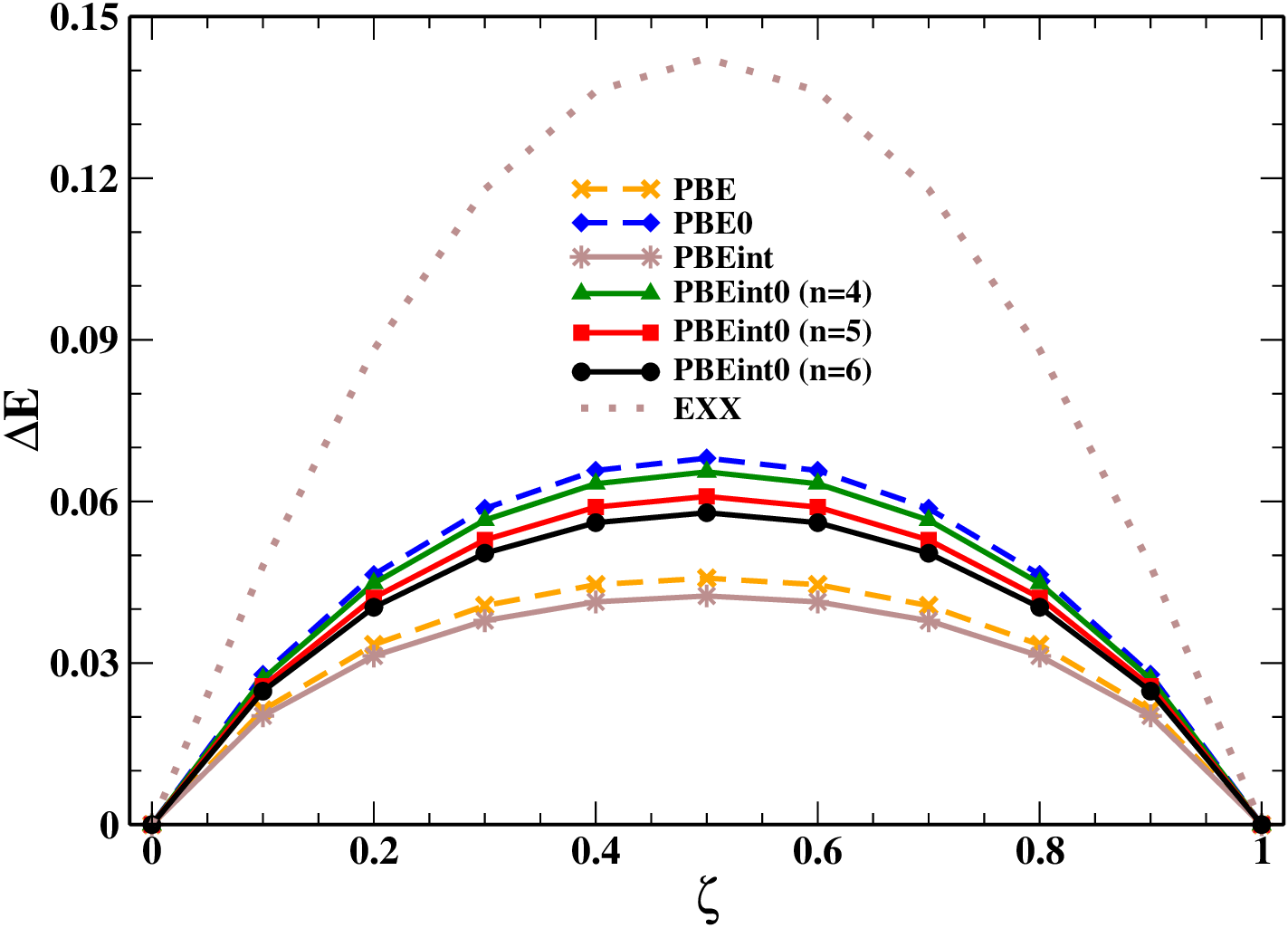}
\caption{ $\Delta E=E[n,\zeta]-E[n,\zeta=1]$ (Hartree) versus
the spin-polarization $\zeta$.}
\label{f2}
\end{figure}

For the static correlation case, the exact exchange 
fails badly, and thus the global hybrids usually increase 
the error in the static correlation with respect to the 
original GGA. So, from this point of view, the HF mixing
should be as small as possible. In Fig. \ref{f2}, we show 
that indeed for the different variants of hPBEint 
increasingly good results are obtained in the series $n$=4, 5, 6.
Thus, hPBEint with $n=6$ has the best performance, while the
worst results are obtained with PBE0, which is slightly worst
than hPBEint with $n$=4.  

We recall however that the 
$N$-electron self-interaction error \cite{PPLB}, 
or the delocalization error \cite{CSY}, that is related to the convexity 
behavior of the functional at fractional number of electrons, 
decreases when the mixing parameter of a global hybrid increases.
Thus, the errors of static correlation and of delocalization can 
not be both reduced by a global hybrid.
%
%

\begin{table*}
 \begin{center}
 \caption{\label{tab_rmae}Mean absolute errors relative to PBEint
 (RMAEs; see Eq. (\ref{e10})) for different sets of tests and all
 functionals. In the last lines are reported: the global RMAE-mol for
 molecular properties (organic molecules, non-bonded interactions,
 transition-metal systems, dipoles); the global RMAE-other for properties
 not concerning molecular systems (interfaces, solid-state systems, model
 systems), in this case for hybrid functionals solid-state RMAEs are not
 considered; the global RMAE-tot for all properties. For each row the
 best results are highlighted with bold style, the worst ones are underlined }
 \begin{ruledtabular}
 \begin{tabular}{lrrrrrrrr}
 Set      & PBE & PBEint & PBEsol & PBE0 & \multicolumn{3}{c}{hPBEint}
 \\
         & & & & & $n=4$ & $n=5$ & $n=6$ \\
 \hline
 Organic molecules & 0.76 & 1.00 & \it{1.21} & \bf{0.60} & 0.79 & 0.78
 &      0.79 \\
 Non-bonded inter. & \bf{0.81} & 1.00 & \it{2.84} & 0.93 & 1.15 & 1.11
 &      1.09 \\
 Transition metals & 1.01 & \bf{1.00} & 1.35 & \it{1.84} & 1.41 & 1.24
 &      1.09 \\
 Dipoles           & 0.99 & 1.00 & \it{1.01} & \bf{0.76} & 0.77 & 0.80
 &      0.82 \\
                   &      &      &           &           &      &      &
 \\
 RMAE-mol          & \bf{0.85} &        1.00 & \it{1.50} & 1.03 & 1.04 & 0.97
 &      0.93 \\
 \hline
 Interfaces        & 1.09 & 1.00 & 1.19 & \it{1.30} & 1.03 & 0.98 &
 \bf{0.93} \\
 Solid-state       & \it{1.46} &        \bf{1.00} & 1.09 & - & - & - & - \\
 Model systems     & 1.12 & 1.00 & 1.05 & \it{1.18} & 1.02 & 0.96 &
 \bf{0.93} \\
                   &      &      &           &           &      &      &
 \\
 RMAE-other        & 1.22 & 1.00 & 1.11 & \it{1.24}$^a$ & 1.03$^a$ &
 0.97$^a$ & \bf{0.93}$^a$ \\
 \hline
 RMAE-tot          & 1.03 & 1.00 & \it{1.39} & 1.10$^a$ & 1.03$^a$ &
 0.98$^a$ & \bf{0.94}$^a$ \\
 \end{tabular}
 \end{ruledtabular}
 \end{center}
 a) It does not include solid-state.
 \end{table*}

\subsection{ Global assessment}

To compare the MAE of different tests, we computed, for each group
of tests, the MAE relative to PBEint defined as
\begin{equation}\label{e10}
RMAE(\mathrm{method}) \equiv \sum_i \frac{MAE_i(\mathrm{method})}{MAE_i(\mathrm{PBEint})},
\end{equation}
where $i$ runs over a given set of tests.
The RMAEs for the different classes of tests considered in this work 
are collected in Tab. \ref{tab_rmae}.
To help the analysis of the results, the benchmarks were 
divided in two subsets:
the {\em molecular-based} tests including organic molecules, 
non-bonded interactions, transition metals and dipoles 
and the {\em other-systems} including interfaces, 
solid-state and model systems.
For the two groups RMAE-mol and RMAE-other indicate respectively 
the average RMAE.
Finally, RMAE-tot indicates the averaged RMAE among all seven benchmarks. 

Among the considered GGAs, PBEint shows a remarkably good performance
yielding the best global RMAE-other and RMAE-tot and a 
RMAE-mol not much higher than PBE and even lower than PBE0. 
In fact, despite PBEint provides the smallest MAE only in 
few cases, it has a more balanced performance on different tests.
On the contrary, PBE0 and PBE give very accurate results in some cases but
behave rather poorly for other tests. 
The good balance over a broad set of systems and properties is an 
important feature of the PBEint functional which makes it 
suitable for applications in complex systems where different 
situations (or density regimes) may coexist. 

Concerning the hybrid functionals similar considerations apply.
We observe that functionals incorporating  25\% of Hartree-Fock 
exchange (PBE0 and hPBEint with $n=4$),
despite working very well for energetic properties of
organic molecules (see Table \ref{tab1}), perform 
globally worst than their GGA counterparts and other hybrids with
a lower Hartree-Fock content, again
because of important failures for some properties (e.g., structures and
reaction energies). The best overall results are obtained with hPBEint
using $n=6$, which gives also a RMAE-mol of 0.95,
(better than PBEint and PBE0). 
Notably, hPBEint $n=6$ has in fact a nicely uniform
performance for all the tests considered in the present work, being
never the worst one and showing a good accuracy for
all systems and properties. 

Moreover, hPBEint with $n=6$ and PBEint are
the only functionals that can describe rather accurately at the same time
energetic and structural properties (this holds also for bulk in the case 
of PBEint).

\section{Conclusions}
In this paper, we performed an extended study of the 
PBEint global hybrids. We tested important energetical and structural
properties of organic molecules and metal complexes, as well as
metal-molecule interfaces. We showed that the hPBEint functional,
with the optimal mixing parameter $n$=6, maintains the well-balanced
behavior of the PBEint GGA functional, improving the
overall accuracy, over a wide range of problems
and can thus be successfully employed in complex studies such as the 
description of hybrid metal-molecule interfaces.

In fact, the PBEint functional and its hybrid forms 
take benefit from the good treatment of both the slowly- and 
rapidly-varying density regimes, granted by Eq. (\ref{e4}).
Therefore, despite some limitations for the
thermochemistry of organic molecules, 
they provide reasonably accurate results for all the tests considered
and outperform the other functionals for transition metals,
hybrid interfaces and solid-state systems.
For these reasons the PBEint and hPBEint($n$=6) functional are
especially appealing for applications involving transition-metal
clusters and/or the interaction of metal clusters with organic
molecules. On the other hand, because Eq. (\ref{e4}) is based on a
simple GGA ansatz, PBEint-based functionals cannot be expected to 
outperform specialized functionals for specific problems
(e.g. atomization energies of small organic molecules).

In the case of hPBEint the good behavior of the underlying PBEint 
GGA allows to achieve a good compromise in the overall performance using
only a small fraction of Hartree-Fock exchange (16.67\% for $n$=6), 
similarly with the case of hybrid meta-GGAs.
This suggests that the PBEint GGA functional could be an interesting 
starting point for the construction of more advanced hybrids 
(e.g. local hybrids or screened ones) of good accuracy and large 
applicability in computational studies.

Finally, we note that 
a recent work \cite{zeta} showed that a GGA can be accurate for both 
solids and molecules, if it satisfies a statistical constraint for 
atomization energies derived from simple one-electron densitites. 
In particular, the zPBEsol and zPBEint functionals 
of Ref. \cite{zeta} significantly improve the 
atomization energies and other spin-dependent properties, performing 
the same as the original functionals for any closed-shell system. Thus, 
they can be a good starting point for future developments of the
hPBEint functional.

\section{Acknowledgments} We thank TURBOMOLE GmbH for providing us 
the TURBOMOLE program package  and M. Margarito for technical support. 
This work was funded by the ERC Starting Grant FP7 Project DEDOM, 
Grant Agreement No. 207441.

\end{document}